%% file: kuang2016.tex
\documentclass{sig-alternate-05-2015}

\usepackage[breaklinks]{hyperref}
\usepackage{natbib}
\usepackage{graphicx,import}
\usepackage{transparent}

\usepackage{caption}
\usepackage{capt-of}

\usepackage{bm}

\usepackage{booktabs}
\usepackage{siunitx}

\usepackage{subfigure}

\usepackage{array}
\newcolumntype{L}[1]{>{\raggedright\let\newline\\\arraybackslash\hspace{0pt}}m{#1}}
\newcolumntype{C}[1]{>{\centering\let\newline\\\arraybackslash\hspace{0pt}}m{#1}}
\newcolumntype{R}[1]{>{\raggedleft\let\newline\\\arraybackslash\hspace{0pt}}m{#1}}

\usepackage{color, colortbl}
\definecolor{LightGray}{gray}{0.9}
\definecolor{DeepGray}{gray}{0.7}
\definecolor{Red}{rgb}{0.855,0.333,0.153}
\definecolor{Blue}{rgb}{0.039,0.451,0.725}
\definecolor{Green}{rgb}{0.196,0.753,0.478}

\usepackage{enumitem}

\newenvironment{prettyitem}[1]{
\begin{itemize}[topsep=0pt,leftmargin=#1]
\setlength\itemsep{-0.5em}}{
\end{itemize}}

\begin{document}

\setcopyright{acmcopyright}

\doi{}

\isbn{}

\conferenceinfo{KDD '16}{August 13--17, 2016, San Francisco, CA, USA}

\acmPrice{\$15.00}

%

\title{Computational Drug Repositioning Using \\ Continuous Self-controlled Case Series} 

%
%
%
%

\numberofauthors{6} 
%

\author{Zhaobin Kuang\textsuperscript{1}, James Thomson\textsuperscript{2}, Michael Caldwell\textsuperscript{3},\\Peggy Peissig\textsuperscript{4}, Ron Stewart\textsuperscript{5}, David Page\textsuperscript{6}\\
\email{UW-Madison\textsuperscript{1,6}, Morgridge Institute\textsuperscript{2,5}, Marshfield Clinic\textsuperscript{3,4}}\\
\email{zkuang@wisc.edu\textsuperscript{1}, page@biostat.wisc.edu\textsuperscript{6}},\\
\email{JThomson@morgridge.org\textsuperscript{2}, 
RStewart@morgridgeinstitute.org\textsuperscript{5},}\\
\email{caldwell.michael@marshfieldclinic.org\textsuperscript{3},
Peissig.Peggy@mcrf.mfldclin.edu\textsuperscript{4}}}

\maketitle
\begin{abstract}
Computational Drug Repositioning (CDR) is the task of discovering potential new indications for existing drugs by mining large-scale heterogeneous  drug-related data sources. Leveraging the patient-level temporal ordering information between numeric physiological measurements and various drug prescriptions provided in Electronic Health Records (EHRs), we propose a Continuous Self-controlled Case Series (CSCCS) model for CDR. As an initial evaluation, we look for drugs that can control Fasting Blood Glucose (FBG) level in our experiments. Applying CSCCS to the Marshfield Clinic EHR, well-known drugs that are indicated for controlling blood glucose level are rediscovered. Furthermore, some drugs with recent literature support for the potential effect of blood glucose level control are also identified.
\end{abstract}

%
%

\ccsdesc[500]{Mathematics of computing~Regression analysis}
\ccsdesc[500]{Applied computing~Health care information systems}
\ccsdesc[500]{Applied computing~Health informatics}

%
%

%
%
\printccsdesc

\vspace{-2mm}
\keywords{Longitudinal Data; Self-Controlled Case Series; Computational Drug Repositioning}

\vspace{-1mm}
\section{Introduction}
Drug repositioning is the task of identifying new potential indications for existing drugs. This task has been steadily rising to prominence because the traditional process of \textit{de novo} drug discovery can be slow, expensive, and risky \citep{ashburndrug2004}. Moreover, with the advent of the big data era, abundant data sources that collect rich drug-related information are emerging.  Mining large-scale heterogeneous drug-related data sources,  Computational Drug Repositioning (CDR) has become an active research area that has the potential to deliver more effective drug repositioning. There have been several comprehensive reviews in the literature on  CDR \citep{hurlecomputational2013,lia2015}. Many methods leverage genotypic and transcriptomic information \citep{lambthe2007,kuhna2010}, as well as drug molecular structure and drug combination information \citep{liudcdb:2010,knoxdrugbank2011}. A prior study that used Electronic Health Records (EHRs) to validate a potential indication of \textit{one} existing drug has also been reported \citep{xuvalidating2014}.

We are interested in mining EHRs in order to identify  a potential indication from \textit{multiple} existing drugs simultaneously. As an initial attempt, we examine the \textit{numeric} values of fasting blood glucose (FBG) level recorded in patients' EHRs \textit{before} and \textit{after} some drugs are prescribed to those patients, in the hope of identifying previously unknown potential uses of drugs to control blood glucose level. 

For this purpose, we extend the Self-Controlled Case Series (SCCS) \citep{simpsonmultiple2013} model that has been widely used in the Adverse Drug Reactions (ADRs) discovery community to handle \textit{continuous} numeric response, hence the name of our model, Continuous Self-Controlled Case Series (CSCCS).

The cornerstone of a self-controlled method is an understanding of how drug prescription history will potentially influence the FBG level every time such a measurement is taken. For example, an antibiotic drug taken ten years ago might have less, if any, influence on the FBG level than an anti-diabetic drug taken a day before that FBG level is measured. To determine how long a drug can potentially influence a patient, we furthermore propose a data-driven approach that leverages change point detection \citep{muggeoestimating2003}, resulting in estimations of different time spans of influence for different drugs.

Our contributions are three-fold:
\begin{prettyitem}{0pt}
\item To the best of our knowledge, this is the first translation of SCCS methodology from ADR discovery to CDR. Our work is a pilot study evaluating the use of temporal ordering information between numeric physical measurements and drug prescriptions available in EHRs for the knowledge discovery process of CDR.
\item Based on the insightful observations of \cite{xuuse2012}, we derive our CSCCS model from a fixed effect model and hence extend the original SCCS model to address continuous numeric response variables.
 \item We introduce to the CDR and ADR discovery community a data-driven approach for adaptively determining the time spans of influence of different drugs to the patients.
\end{prettyitem}
 
\section{Continuous Self-controlled Case Series (CSCCS) Model}
\begin{figure}
\tiny\centering
\scalebox{1}{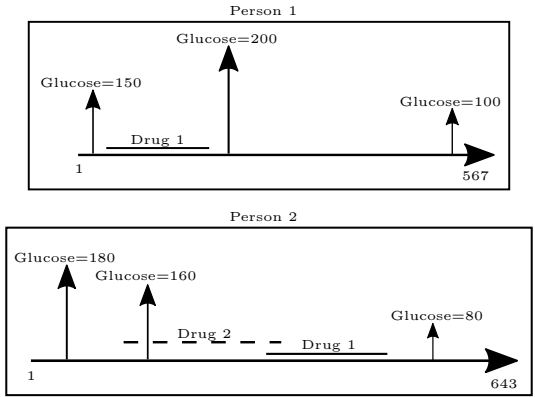}
\caption{An example of EHRs}
\label{fig:ehr}\vspace{-4mm}
\end{figure}

\subsection{Notation}
Figure~\ref{fig:ehr} visualizes an example of health records for two patients. To confine the time span of a drug that has potential influence on that patient, we use the concept of \textit{drug era}, which is recorded with its start date, end date and the name (or id) of the drug. We consider a patient to be under consistent influence of a drug during a drug era of that drug.  However, drug era information is not readily available in most EHRs. Instead, drug prescription information with the name of a drug and the start date of the prescription is usually provided in observational data. How to construct drug eras from prescription records is a challenging and significant task for both CDR and ADR discovery \citep{nadkarnidrug2010,ryanestablishing}. We provide a data-driven approach to this task in Section~\ref{sec:bp}.

Measurements of FBG level might also be taken from time to time and are recorded with the date taken, as well as their numeric measurement values. We assume that at most one FBG measurement is taken for a particular patient on a particular day.

Let there be $N$ patients with FBG measurements and $M$ different drugs in the EHR. We construct a cohort using all the FBG measurement records as well as all the drug era records from all the $N$ patients. Furthermore, we use a continuous random variable $y_{ij}$, where $i \in \left\{ 1, 2, \cdots, N\right\}$, $j \in \left\{ 1,2,\cdots, J_i \right\}$, to denote the value of the $j^{th}$ FBG measurement taken among a total number of $J_i$ measurements during the observation period of the $i^{th}$ person. Similarly, we use a binary variable $x_{ijm},\ i \in \left\{ 1, 2, \cdots, N\right\}$, $j \in \left\{ 1,2,\cdots, J_i \right\}$, $m \in \left\{1,2,\cdots,M\right\}$ to denote the exposure status of the $m^{th}$ drug of the $i^{th}$ person at the date when the $j^{th}$ FBG measurement is taken, with $1$ representing exposure and $0$ otherwise.

\subsection{The Linear Fixed Effect Model}
We treat the $y_{ij}$'s as the response variables and first consider the following linear regression model:
\begin{equation}
\label{eq:linReg}
y_{ij} \lvert \bm{x}_{ij} = \alpha_i + \bm{\beta}^{\top} \bm{x}_{ij} + \epsilon_{ij},\quad 
\epsilon_{ij} \stackrel{iid}{\sim} N \left(0,\sigma^2\right),
\end{equation}
where
\begin{equation*}
 \bm{\beta} = \left[ \begin{matrix}
 \beta_1 & \beta_2 & \cdots & \beta_M 
\end{matrix} \right]^{\top},\quad 
\bm{x}_{ij} = \left[ \begin{matrix}
 x_{ij1}& x_{ij2}& \cdots &x_{ijM} 
\end{matrix} \right]^{\top},
\end{equation*}
$\alpha_i$, which is called the \textit{nuisance parameter}, represents the individual effect of the $i^{th}$ person on the value of $y_{ij}$, invariant to day $j$, drug $m$, and other patients, and $\epsilon_{ij}$'s are independent and identically distributed Gaussian noises with zero mean and fixed but unknown variance $\sigma^2$.

The parameter of interest in this problem is $\bm{\beta}$, which represents the effect of each of the $M$ drugs on the response $\bm{y}$ when a patient is under the joint exposure statuses specified by $\bm{x}_{ij}$. More specifically, suppose the $m^{th}$ component of $\bm{\beta}$, $\beta_m$, is evaluated to a negative number, that is to say,  exposure to the $m^{th}$ drug will cause the FBG level to decrease. If this drug is not known to be prescribed for lowering FBG, such a decrease is an indicator that this drug might have the potential to be repositioned to help diabetic patients control their blood glucose level, given further investigation. 

In this setting, fitting a linear regression model is equivalent to solving the following least squares problem:
\begin{equation}
\label{eq:fix}
{\arg\min}_{\bm{\alpha},\bm{\beta}} \frac{1}{2} \left\lVert \bm{y} - 
\begin{bmatrix}
\bm{Z} & \bm{X}
\end{bmatrix}
\begin{bmatrix}
\bm{\alpha} \\ \bm{\beta}
\end{bmatrix}
\right\rVert_2^2,
\end{equation}
where
\begin{gather*}
\bm{\alpha} = \begin{bmatrix}
\alpha_1 & \alpha_2 & \cdots & \alpha_N 
\end{bmatrix}^{\top},\quad 
\bm{Z} = \text{diag}\left(\bm{1}_1, \cdots, \bm{1}_N\right),\\
\bm{y} = \begin{bmatrix}
y_{11} & \cdots &  y_{1J_1} & \cdots & y_{N1} & \cdots & y_{NJ_N}
\end{bmatrix}^{\top},\\
\bm{X} = \begin{bmatrix}
\bm{x}_{11} & \cdots & \bm{x}_{1J_1} & \cdots & \bm{x}_{N1} & \cdots & \bm{x}_{NJ_N}\end{bmatrix}^{\top},
\end{gather*}
where $\bm{Z}$ is a block diagonal matrix with $\bm{1}_i$ being a $J_i\times 1$ vector where all the components are $1$.
The least squares problem in (\ref{eq:fix}) is a linear \textit{fixed effect model} with $\bm{\alpha}$ being a nonrandom quantity whose $i^{th}$ component $\alpha_i$,  can be interpreted as the \textit{average} FBG measurement level of the $i^{th}$ patient taken over time without exposing to any drugs.

\subsection{Deriving the CSCCS Model from the Linear Fixed Effect Model}

Like the SCCS model, the motivation behind the CSCCS model is to use only $\bm{\beta}$ as a parsimonious parameterization to predict the response vector $\bm{y}$. Inspired by the work in \citep{xuuse2012}, where the equivalence between the Poisson fixed effect model and the SCCS model is established, we are able to derive the CSCCS model from the linear fixed effect model in (\ref{eq:fix}) in a similar fashion. Let
\begin{equation*}
\ell\left(\bm{\alpha},\bm{\beta}\right) = 
\frac{1}{2} \left\lVert \bm{y} - 
\begin{bmatrix}
\bm{Z} & \bm{X}
\end{bmatrix}
\begin{bmatrix}
\bm{\alpha} \\ \bm{\beta}
\end{bmatrix}
\right\rVert_2^2.
\end{equation*}
We consider,
\begin{equation}
\label{eq:alpha}
\begin{split}
\frac{\partial \ell\left(\bm{\alpha},\bm{\beta}\right)}{\partial \bm{\alpha}} = \bm{0} \Rightarrow \bm{\alpha} = \left(\bm{Z}^{\top}\bm{Z}\right)^{-1}\bm{Z}^{\top}\left(\bm{y}-\bm{X}\bm{\beta}\right) = \bar{\bm{y}} - \bar{\bm{X}}\bm{\beta},
\end{split}
\end{equation}
where $\bar{\bm{y}}$ is an $N\times 1$ vector with the $i^{th}$ component, $\bar{y}_i = \frac{1}{J_i}\sum_{j=1}^{J_i} y_{ij}$, and $\bar{\bm{X}}$ is an $N \times M$ matrix with the $i^{th}$ row, $\bar{\bm{X}}_{i\cdot} = \frac{1}{J_i}\sum_{j=1}^{J_i} \bm{x}_{ij}^{\top}$. Substitute (\ref{eq:alpha}) into (\ref{eq:fix}) results in the CSCCS model:
\begin{equation}
\label{eq:csccs}
{\arg\min}_{\bm{\beta}} \frac{1}{2} \lVert \bm{y} - \bm{Z}\bm{\bar{y}} - \left(\bm{X} - \bm{Z}\bar{\bm{X}} \right) \bm{\beta}\rVert_2^2.
\end{equation}
The model in (\ref{eq:csccs}) is in the desired form of parsimonious parameterization in that the optimization problem is defined only in the sapce of $\bm{\beta}$, and the nuisance parameter $\bm{\alpha}$ is eliminated.

The CSCCS model is a linear model and hence CSCCS is able to predict \textit{continuous} response $\bm{y}$. The model is \textit{self-controlled} in that each FBG measurement and their corresponding drug exposure statuses are adjusted by their mean within each individual. The model also utilizes \textit{case series} in that only cases (patients that have at least one FBG measurement) are admitted in the cohort.

CSCCS is derived from its linear fixed effect model counterpart. This derivation shares the same spirit with the equivalence between the original SCCS and the Poisson fixed effect model; in this sense, CSCCS extends SCCS to address numeric response in the new setting.

Although both models in (\ref{eq:fix}) and (\ref{eq:csccs}) can be considered as  linear  models, from the perspective of implementation efficiency, the explicit form of CSCCS in (\ref{eq:csccs}) is of vital importance  for the task of CDR using large-scale EHRs. This is because the parameter of interest in our task is $\bm{\beta}$ and the nuisance parameters do not provide  direct information in evaluating the impact of a drug in changing FBG level. In the setting of large-scale EHRs, where tens of thousands of patient records might be admitted into the cohort as cases, the dimension of the nuisance parameter can potentially be very high. In this scenario, without the access to a special purpose solver for the fixed effect model, solving a model in the form of (\ref{eq:fix}) using only a general purpose linear model solver can be time consuming or even infeasible. On the contrary,  using the explicit form of CSCCS in (\ref{eq:csccs}), a general purpose linear model solver only needs to find solutions in the space of $\bm{\beta}$, a parameter whose dimension is only as large as the number of drugs available in the cohort, which is a much smaller number than the dimension of nuisance parameters.

\section{Challenges in EHR data}
Several challenges arise when we apply CSCCS to EHR data. In this section, we present the further refinements we perform on the CSCCS model presented in (\ref{eq:csccs}) in order to address these challenges.

\subsection{High Dimensionality}
EHR data is a type of high-dimensional longitudinal data. While tens of thousands of patient records might be admitted into the cohort, effects of thousands of drugs on the FBG level need to be evaluated simultaneously, introducing a high-dimensional problem. This motivates us to incorporate sparsity into our model using the  penalty \citep{tibshirani1996regression},
\begin{equation}
\label{eq:csccsSparse}
{\arg\min}_{\bm{\beta}} \frac{1}{2} \lVert \bm{y} - \bm{Z}\bm{\bar{y}} - \left(  \bm{X} -\bm{Z}\bar{\bm{X}} \right) \bm{\beta}\rVert_2^2 + \lambda \lVert \bm{\beta} \rVert_1,
\end{equation}
where $\lambda>0$ is a tuning parameter determining the level of sparsity. 

The incorporation of this penalty essentially assumes that only a small portion of drugs are related to the change of FBG level, and the rest of them do not have significant effect on changing FBG level when patients are exposed to those drugs. With the $L_1$ penalization, most components of $\bm{\beta}$ will be evaluated to zero or a number that is close to zero. The result is, instead of evaluating the effect of \textit{each} of the $M$ drugs on FBG level, $L_1$ penalized CSCCS only selects a subset of drugs that, in some sense, are most correlated to the change of FBG level, and estimates their relative strength and direction of change among the drugs chosen.

\subsection{Irregular Time Dependency}
\label{sec:csccsa}
The linear fixed effect model assumes that all responses are independent of each other. The meaning of independence is two-fold. On one hand, responses from different patients are independent of each other. To explain differences across patients (e.g.~some patients tend to have higher FBG levels than others in general), $\bm{\alpha}$ is used with each component representing the time-invariant effect of each patient on the response. On the other hand, responses  observed at different time are independent of each other. To explain differences across time (e.g.~FBG levels observed in early age \textit{might be} lower than those in old age), a time-dependent variable that has the same value across all patients can be introduced. That is to say:
\begin{equation}
\label{eq:twoWay}
y_{ij} \lvert \bm{x}_{ij} = \alpha_i + t_j + \bm{\beta}^{\top}\bm{x}_{ij} + \epsilon_{ij},
\end{equation}
where $t_j$ is the time-dependent nuisance parameter whose value depends only on the time when the $j^{th}$ measurement is taken. If observations are recorded regularly across time, (\ref{eq:twoWay}) defines a \textit{two-way fixed effect model}, as opposed to the \textit{one-way fixed effect model} defined in (\ref{eq:fix}) \citep{freeslongitudinal2004}.

In practice, a one-way model might be preferred over a two-way model if we assume that the heterogeneity across different individuals is much more significant than that across time. However, in the task of CDR from EHRs, this assumption might be too restrictive. To begin with, EHRs usually contain observational data of patients that are recorded over decades. Therefore, it is probable that the baseline FBG levels of  patients change significantly over the years. This is especially true when some persistent FBG level altering events, such as the diagnosis of diabetes, occur to some patients. Furthermore, the length of observation periods varies dramatically among patients. Therefore, we do not have a fully observed and consistent dataset to model the set of time-dependent nuisance parameters. Last but not least, the incorporation of time-dependent nuisance parameters is proposed in a setting where data are collected regularly. With the irregular nature of EHR data, modeling time-dependent nuisance parameters directly with a classic two-way fixed effect model is impractical.

To address the aforementioned challenges without much loss in efficiency, we consider a reasonable assumption: given $y_{ij}$ and $y_{ij'}$, where $j \ne j'$, but the dates of the two measurements taken are very close to each other, we assume the two corresponding time-dependent nuisance parameters are equal to each other, i.e.~$t_j = t_{j'}$. More specifically,
\begin{gather*}
y_{ij} \lvert \bm{x}_{ij} = \alpha_i + t_j + \bm{\beta}^{\top} \bm{x}_{ij} + \epsilon_{ij},\\
y_{ij'} \lvert \bm{x}_{ij'} = \alpha_i + t_{j'} + \bm{\beta}^{\top} \bm{x}_{ij'} + \epsilon_{ij'},\\
\lvert d_{ij} - d_{ij'} \rvert \le \tau \Rightarrow t_j = t_{j'},
\end{gather*}
where $d_{ij}$ and $d_{ij'}$ represent that the $j^{th}$ and $j'^{th}$ measurements of the $i^{th}$ patient are taken at the $d_{ij}^{th}$ day and $d_{ij'}^{th}$ day of the observation period, and $\tau$ is a predetermined threshold. Then,
\begin{equation}
\label{eq:diff}
\mathbb{E}\left[ y_{ij} - y_{ij'} \lvert \bm{x}_{ij}, \bm{x}_{ij'}\right] = \bm{\beta}^{\top}\left(\bm{x}_{ij} - \bm{x}_{ij'}\right) \equiv \bm{\beta}^{\top}\bm{\delta}_{ij},
\end{equation}
where the nuisance parameters are eliminated. Therefore, the quantity in (\ref{eq:diff}) depends only on $\bm{\beta}$ and the data. 

Based on this formulation, we can reconstruct the CSCCS model to address irregular time dependency as follow: firstly, given $\tau$, construct a cohort where only patients with at least a pair of FBG measurements taken within $\tau$ days are admitted; only adjacent pairs are used. Secondly, solve the following lasso problem:
\begin{equation}
\label{eq:csccsAdj}
{\arg\min}_{\bm{\beta}} \frac{1}{2} \lVert \bm{D}\bm{y} - \bm{D}\bm{X}\bm{\beta}\rVert_2^2 + \lambda \lVert \bm{\beta} \rVert_1,
\end{equation}
where $\bm{D}$, when multiplied with $\bm{y}$ or $\bm{X}$, generates the difference between the measurement of an earlier record and the corresponding measurement of its adjacent later-measured record of the same patient, with the constraint that the two records are collected within a time span of $\tau$ days. 

Note that the model in (\ref{eq:csccsAdj}) is not equivalent to the model in (\ref{eq:csccsSparse}). However, the model in (\ref{eq:csccsAdj}) can still be considered as a variant of CSCCS in that its parameterization is still restricted to $\bm{\beta}$, with the goal of predicting a continuous response, using data subtraction within the \textit{same} patient as a self-controlled mechanism, and only admitting cases into the cohort. We call the the model in (\ref{eq:csccsAdj}) as CSCCS for Adjacent response, or CSCCSA.

\subsection{Confounding}
Another challenge an algorithm must tackle is the confounding issue arises due to the complex nature of clinical observational data. In the setting of EHRs, one important confounding issue is called \textit{confounding by co-medication}. Consider drug A and drug B, where only drug A can lower FBG level and drug B has no significant effect on changing blood sugar. However, drug B is usually prescribed with drug A. In this case, drug B can be a confounder if we only evaluate the marginal correlation between each drug and FBG level. Another confounding issue in this setting is \textit{confounding by comorbidity}. Consider the FBG-lowering drug A given to a diabetic patient. Following the prescription of drug A, some other conditions could occur to this patient since diabetes can lead to various comorbidities \citep{Manag30:online}. To treat a newly introduced condition, drug B is prescribed to the patient. In this case, if we again consider only the marginal correlation between drug B and FBG level, one might draw the conclusion that drug B could lower FBG level since after the prescription of drug A, the FBG level has decreased.

In the two aforementioned confounding issues, drug B is called an \textit{innocent bystander}. Like multiple SCCS \citep{simpsonmultiple2013}, multiple CSCCS can effectively handle the innocent bystander confounding problem (a.k.a.~Simpson's Paradox). This is because the confounder seems to spuriously correlated to the FBG level when we consider their marginal correlation. However, using a multiple linear model like CSCCS, the joint exposure statuses of both drug A and drug B can be considered simultaneously. Thereofre, CSCCS might be able to identify that the decrease of FBG level occurs only when conditioning on the exposure of drug A and hence rule out drug B in the model.

In terms of addressing various confounding issues, CSCCS inherits most of the strengths and weaknesses from SCCS, due to the close relationship between the two models. While CSCCS might address reasonably well the innocent bystander confounding problem, it might not be well suited to handle confounding issues such as time-varying confounding \citep{danielmethods2013}. In Section~\ref{sec:experiment}, we empirically evaluate the performance of CSCCS in the CDR task and illustrate how its performance is related to its capabilities of addressing various confounding issues.
\begin{figure}[t!]
\centering
\includegraphics[scale=0.5]{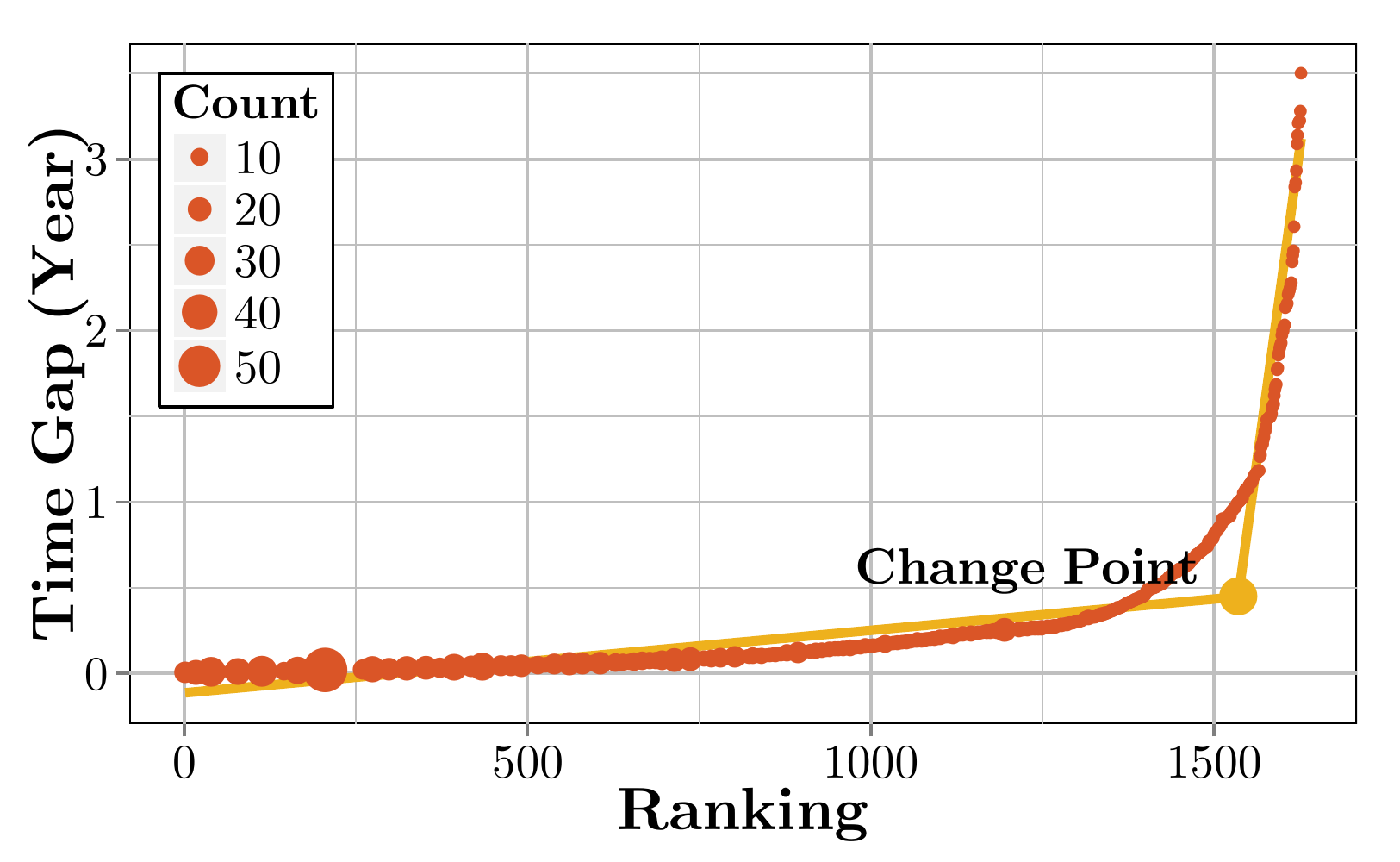}
\caption{Time gap of humalog in asecneding order: the size of dots represents the number of time gaps that share the same value}
\label{fig:humalog}
\end{figure}
\section{A Data-Driven Approach to Construct Drug Eras from Drug Prescription Records}
\label{sec:bp}
A prerequisite of CSCCS is the availability of drug era information of each drug prescribed to each patient. However, drug era information is usually not provided in most EHRs. Instead, drug prescription records of each patient are kept, usually with the name (or id) of the drug and the date of prescription. Constructing drug eras from drug prescription records is an important but challenging task for both CDR using CSCCS and ADR discovery. 

\subsection{Drug Era in Common Data Model}
A heuristic proposed in the Common Data Model (CDM) \citep{reisingerdevelopment2010} by Observational Medical Outcome Partnership (OMOP) is to first consider the prescription dates of each prescription record as the start date of the drug era. It then assumes that each drug era lasts $n$ days and hence computes the end date of the drug era accordingly. Within the same patient, we assume there is only one drug prescription record of the same drug in a given date. In this way, drug eras of the same drug within each patient constructed as before start from different dates. For an adjacent pair of drug eras of the same drug within the same patient, we call the drug era that starts earlier a \textit{former era}, and the other a \textit{latter era}. CDM defines a parameter called \textit{persistence window}. If the start date of the latter era, subtracted by the end date of the former era, is no larger than the persistence window, CDM merges the two drug eras into one, using the start date of the former era as the start date of the new era and the end date of the latter era as the end date of the new era. CDM tries to merge as many drug eras of the same drug within the same patient as possible in this fashion, until every resultant drug era of the same drug within the same patient is separated by more than persistence window amount of time. In CDM, both $n$ and the persistence window are usually set to thirty days.

The intuition behind this heuristic is to build a longer drug era if the prescription date of an adjacent pair of records of the same drug are close enough to each other. A natural question to ask is \textit{how large} the time gap between the two adjacent prescription records can be for us to still consider them close enough?

\begin{figure}[t!]
\centering
\includegraphics[scale=0.5]{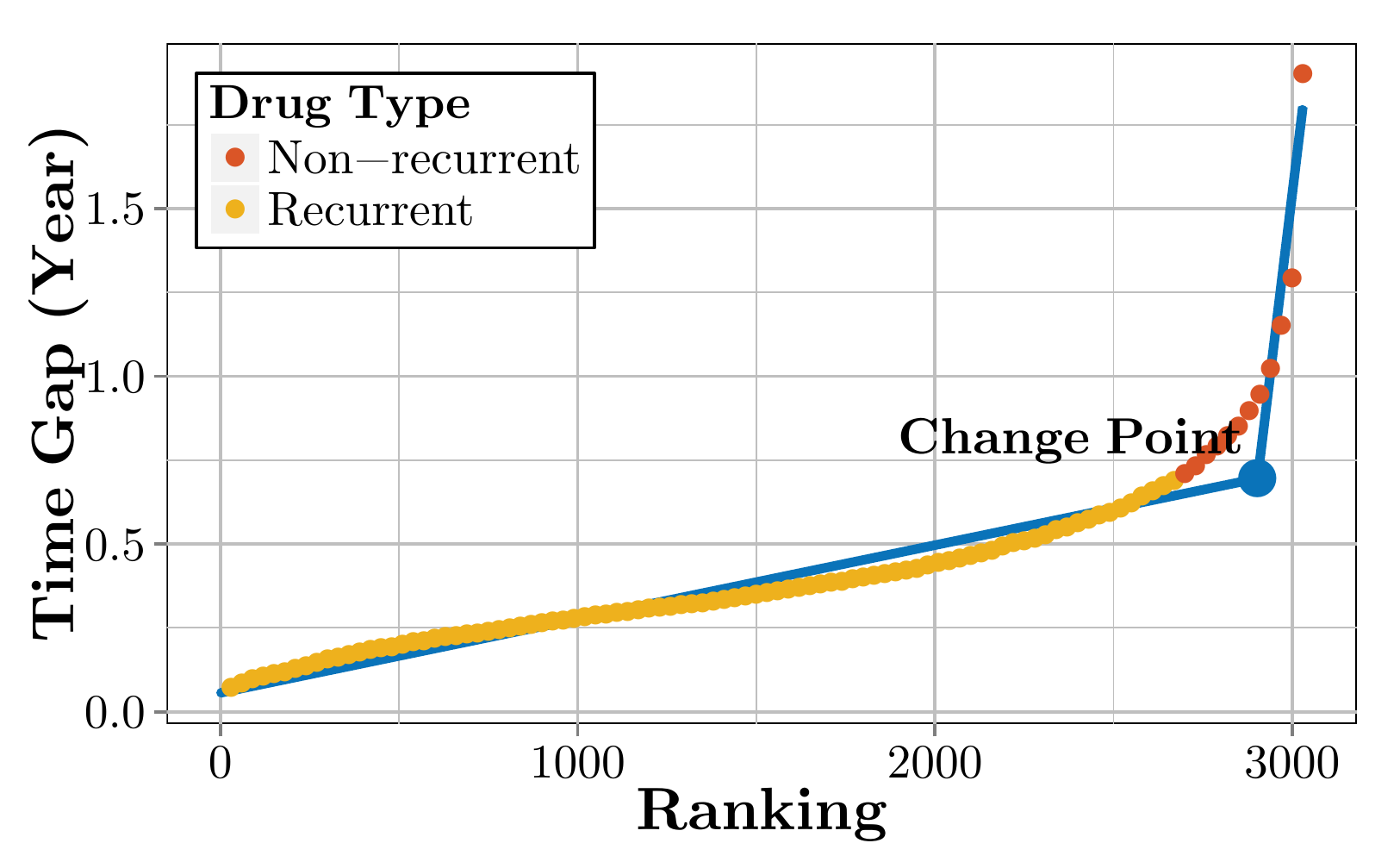}
\caption{Change points of all drugs in the EHRs in ascending order}\vspace{-2mm}
\label{fig:threshold}
\end{figure}
\subsection{Drug Era Construction via Change Point Analysis}
Instead of specifying a predetermined threshold on time gap as it is in CDM, we answer this question via a data-driven approach: for each drug, we compute the time gaps between all adjacent pairs of prescription records. We then sort these time gaps in ascending order. A visualization of the values of the time gaps of Humalog against their relative rankings is given in Figure~\ref{fig:humalog}. From Figure~\ref{fig:humalog}, we notice that the distribution of time gaps can be approximated by a piecewise linear model with a change point close to the end of the sample with large time gap values. The smaller time gaps can be fitted well by the flat linear segment of the model while the larger time gaps can be fitted well by the steep linear segment. This phenomenon leads to a reasonable assumption that the smaller time gaps are sampling from a different underlying distribution than that of the larger time gaps. The smaller time gaps sampling from the same distribution correspond to the adjacent pairs of prescription records that we can consider close enough to each other to construct a lasting drug era. A threshold we can use to distinguish the two types of time gaps is the change point of the piecewise linear model.

For each drug with at least fifty prescription records in the EHRs, we perform change point detection analysis in the aforementioned fashion using \texttt{R} package \texttt{segmented}. We plot the change points of all the drugs against their relative rankings after sorting them in ascending order in Figure~\ref{fig:threshold}. Interestingly, there is also a change point in Figure~\ref{fig:threshold}. A possible explanation of the existence of a change point in Figure~\ref{fig:threshold} is that in EHR data, drug prescriptions of some particular drugs are recurrent in order to battle chronic disease. For example, a diabetic patient needs long-term prescriptions of some FBG lowering drugs. On the other hand, the prescriptions of some other drugs are non-recurrent, such as antibiotics. We consider the change point in Figure~\ref{fig:threshold} as a threshold to distinguish recurrent drugs from non-recurrent drugs in the EHR because a reasonable expectation is that if a drug is recurrent, the gap between an adjacent pair of prescription records of that drug from the same patient will tend not to be too large and hopefully under the change point specified in Figure~\ref{fig:threshold}.

We extend the heuristic provided in CDM as follow:  We first denote the mean of all change point values of the recurrent drugs in the EHR as $\gamma$. For all the recurrent drugs, we set their corresponding $n$'s and the value of their persistence windows to $\frac{\gamma}{2}$. We then set $n=0.04 \text{year}$ (approximately two weeks) for all non-recurrent drugs and $0$ as the value of their persistence windows. 
\begin{figure*}[t!]
\subfigure[PM, CSCCS, and CSCCSA]{\includegraphics[width=0.33\textwidth]{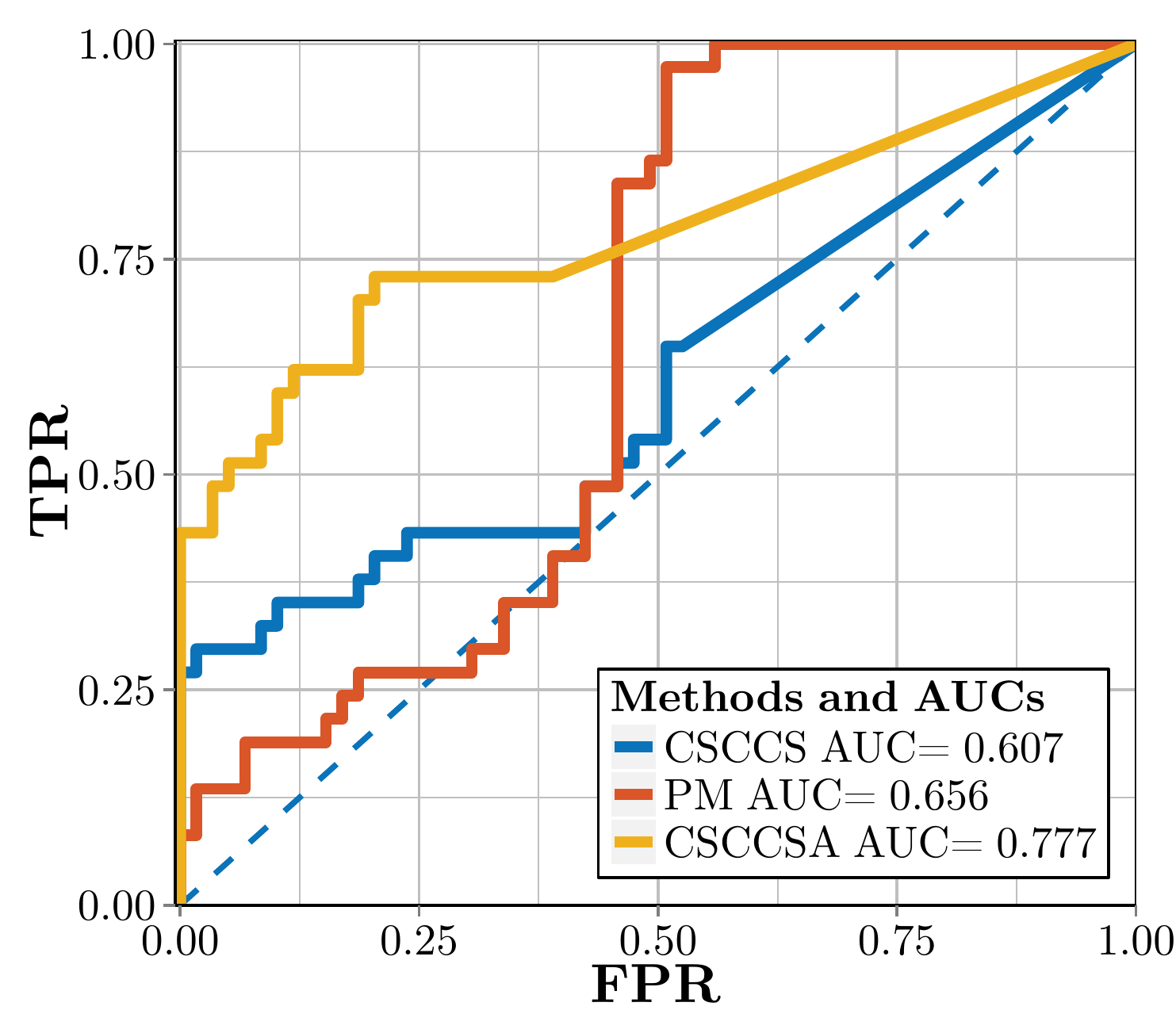}\label{fig:all}}
\subfigure[CSCCS]{\includegraphics[width=0.33\textwidth]{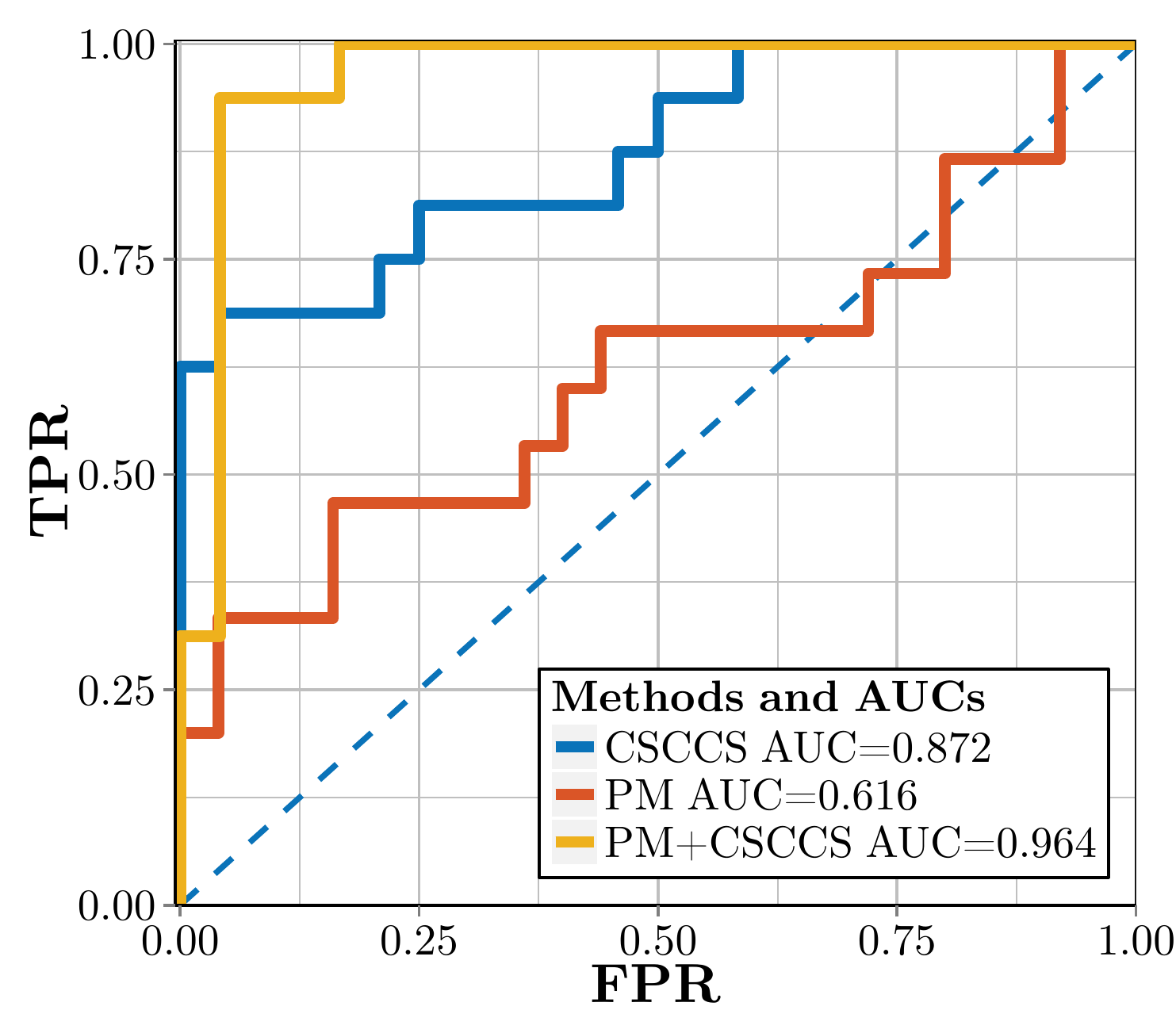}\label{fig:csccs}}
\subfigure[CSCCSA]{\includegraphics[width=0.33\textwidth]{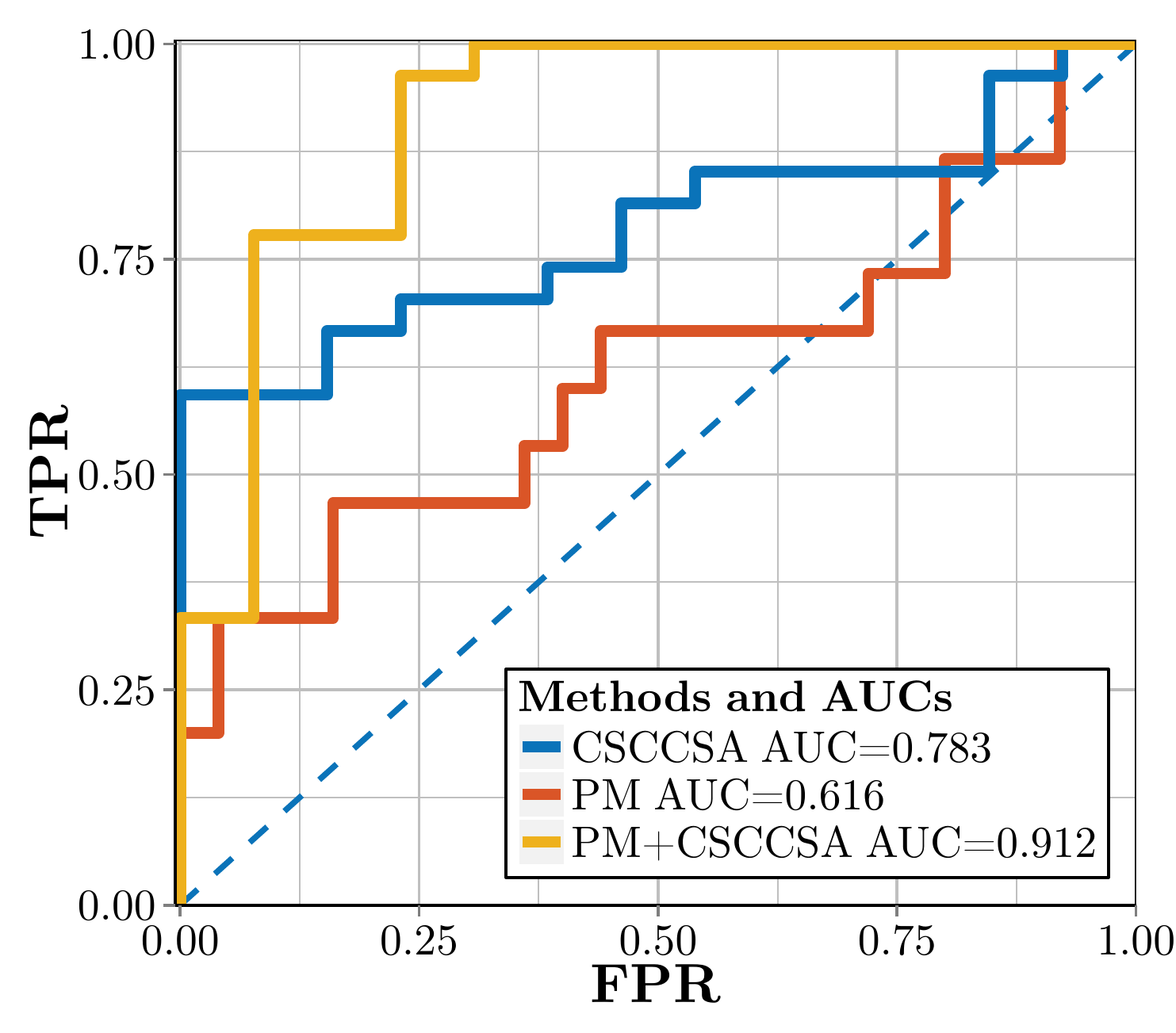}\label{fig:csccsa}}
\vspace{-4mm}
\caption{ROC curves}
\vspace*{-4mm}
\end{figure*}

\section{Experiments}
\label{sec:experiment}
As far as we know, our CSCCS model is the first of its kind to explicitly use temporal ordering information in EHRs for CDR. How do we evaluate the performance of a method that utilizes this type of information? As a preliminary endeavor, we try to answer this question by addressing two major challenges for our experiments.
\subsection{Lack of a Baseline Method}
\label{sec:baseline}
The first challenge we need to handle is the lack of a baseline method that also utilizes temporal ordering information in an EHR for CDR. Inspired by the idea of disproportionality analysis from the pharmacovigilance literature \citep{montastrucbenefits2011}, we propose the \textit{Pairwise Mean} (PM) method as a baseline method. PM assigns a real-valued score to each of the $M$ drugs in the EHR to represent how likely the drug decreases FBG level, and a smaller score implies a stronger decreasing tendency. The score of the $m^{th}$ drug, $s_m$, is computed as follow: first, for the $i^{th}$ patient who has FBG measurements within two years before \textit{and} after the \textit{first} prescription of the $m^{th}$ drug, we compute the mean of those FBG measurements before and after the first prescription, denoted as $b_{mi}$ and $a_{mi}$, respectively; second, compute $s_m$ as:
\begin{equation*}
s_m = \frac{1}{N_m}\sum_{i=1}^{N_m} \left(a_{mi} - b_{mi}\right),
\end{equation*} 
where $N_m$ is the number of patients that have FBG measurements two years before and after the first prescription of the $m^{th}$ drug.

\subsection{Incomplete Ground Truth}
Unlike the task of ADR discovery from the EHR, where numerous research efforts have been invested on developing a  set of ground truth \citep{omop.6:online} drug-adverse-reaction pairs so that algorithms can be run and evaluated, we do not have access to such a  ground truth set for the task of CDR from EHRs. We use Marshfield Clinic EHR as our data source and there are about two thousand drugs for evaluation. To evaluate the performance of our algorithm without knowing the glucose altering effect of every drug, we focus on the top forty most promising drugs generated by PM, CSCCS, and CSCCSA, as shown in Table~\ref{tab:pm}, Table~\ref{tab:csccs}, and Table~\ref{tab:csccsa}, respectively.

In these three tables, rows that are shaded in green represent the drugs commonly prescribed for lowering glucose  while rows that are shaded in red represent the drugs commonly prescribed for increasing glucose. The two types of drugs in the three tables are all manually labeled. Drugs in the unshaded rows might potentially be irrelevant, or might constitute new discoveries. These drugs are discussed in further detail in Section~\ref{sec:potential}. A summary of the number of each of the three types of drugs discovered by the three algorithms are given in Table~\ref{tab:summary}.

In CSCCSA, we set $\tau$ defined in Section~\ref{sec:csccsa} to four years. In Table~\ref{tab:pm}, the counts and scores are $N_m$'s and $s_m$'s defined in Section~\ref{sec:baseline}, while in Table~\ref{tab:csccs} and Table~\ref{tab:csccsa}, the counts are the $L_1$ norm of the columns in $\bm{X}$  corresponding to different drugs, and the scores are the regression coefficients of different drugs. We only consider drugs with counts greater than or equal to eight. For CSCCS and CSCCSA, we first used a lasso penalty for variable selection to generate a long list of about two hundred drugs, and we present the top forty among those selected drugs as the short list. The number eight and forty could be tuned to optimize accuracy but were fixed here beforehand for practical reasons. Drugs with fewer than eight prescriptions might not have sufficient evidence to support a new use. Evaluating more than forty results per method was too large a burden for human literature review.
\begin{center}\vspace{-2mm}
\captionof{table}{A summary of three types of drugs discovered by the three algorithms}
\begin{tabular}{cccc}
\hline
& PM & CSCCS & CSCCSA \\
\hline
decrease & 15 & 16 & 27 \\
increase & 1 & 1 & 0 \\
potential & 24 & 23 & 13 \\
\hline
\end{tabular}
\label{tab:summary}
\end{center}

\subsection{Receiver Operating Characteristic}
\label{sec:roc}
As shown in Tables~\ref{tab:pm}--\ref{tab:csccsa}, all three methods capture a reasonable number of drugs that are prescribed for lowering glucose among their top forty candidates. We therefore consider identifying drugs prescribed for glucose-lowering as a binary classification task and use Receiver Operating Characteristics (ROC) curves as well as Area Under ROC (AUROC) to evaluate the performance of each algorithm. 

We first construct the ROC curves of the three methods using the union list of drugs from Tables~\ref{tab:pm}--\ref{tab:csccsa}. The three ROC curves are presented in Figure~\ref{fig:all}. Since we perform variable selection in CSCCS and CSCCSA, some drugs might be assigned scores of zero and hence are considered irrelevant to the prediction of FBG level. In these cases, we put these drugs at the bottom of the union list and consider them to be identified as positive examples by the algorithms only at the very end. This results in the straight line segment of the ROC curves of CSCCS and CSCCSA at the liberal region. Figure~\ref{fig:all} shows that CSCCSA has the highest AUC, outperforming CSCCS and PM by a significant margin, while PM and CSCCS have similar AUCs. However, in the more conservative region where there is drug support for all three methods, CSCCS outperforms PM while CSCCSA maintains the best performance. This phenomenon suggests that the modeling assumptions of CSCCS and CSCCSA  are able to provide insights into making reasonable prediction of FBG level.

Figure~\ref{fig:csccs} uses the forty drugs in Tables~\ref{tab:pm} and \ref{tab:csccs} to generate the ROC curves,  in red for PM and in blue for CSCCS. As a comparison, we also plot the ROC curve of the following ensemble strategy: we first use the top forty drugs in Table~\ref{tab:csccs} as a result of variable selection via CSCCS, then we compute the PM scores over the selected drugs. By comparing the AUCs of the three curves, we notice that the ensemble method outperforms CSCCS and PM, while CSCCS outperforms PM. Since the scores used to construct the CSCCS ROC curve are regression coefficients of drug exposure statuses under a lasso penalty, the lack of an oracle property for the lasso \citep{wainwrightsharp2006} might potentially trade off the inherent order among drugs for a sparse model. However, such a trade-off is beneficial, based on the significant improvement of AUC of CSCCS compared with the AUC of PM.

Figure~\ref{fig:csccsa} is generated similarly as Figure~\ref{fig:csccs}. The ensemble of CSCCSA with PM outperforms the two individual algorithms. Although the AUC of CSCCSA is less than that of CSCCS, it is worthy to notice that all but one true positive drugs in Table~\ref{tab:csccs} are discovered in Table~\ref{tab:csccsa} at the top fifteen positions. Other than that, CSCCSA is also able to discover twelve more true positives that CSCCS does not capture among its top forty discoveries.

\subsection{Precision at K}
\begin{figure}[t!]
\centering
\includegraphics[scale=0.5]{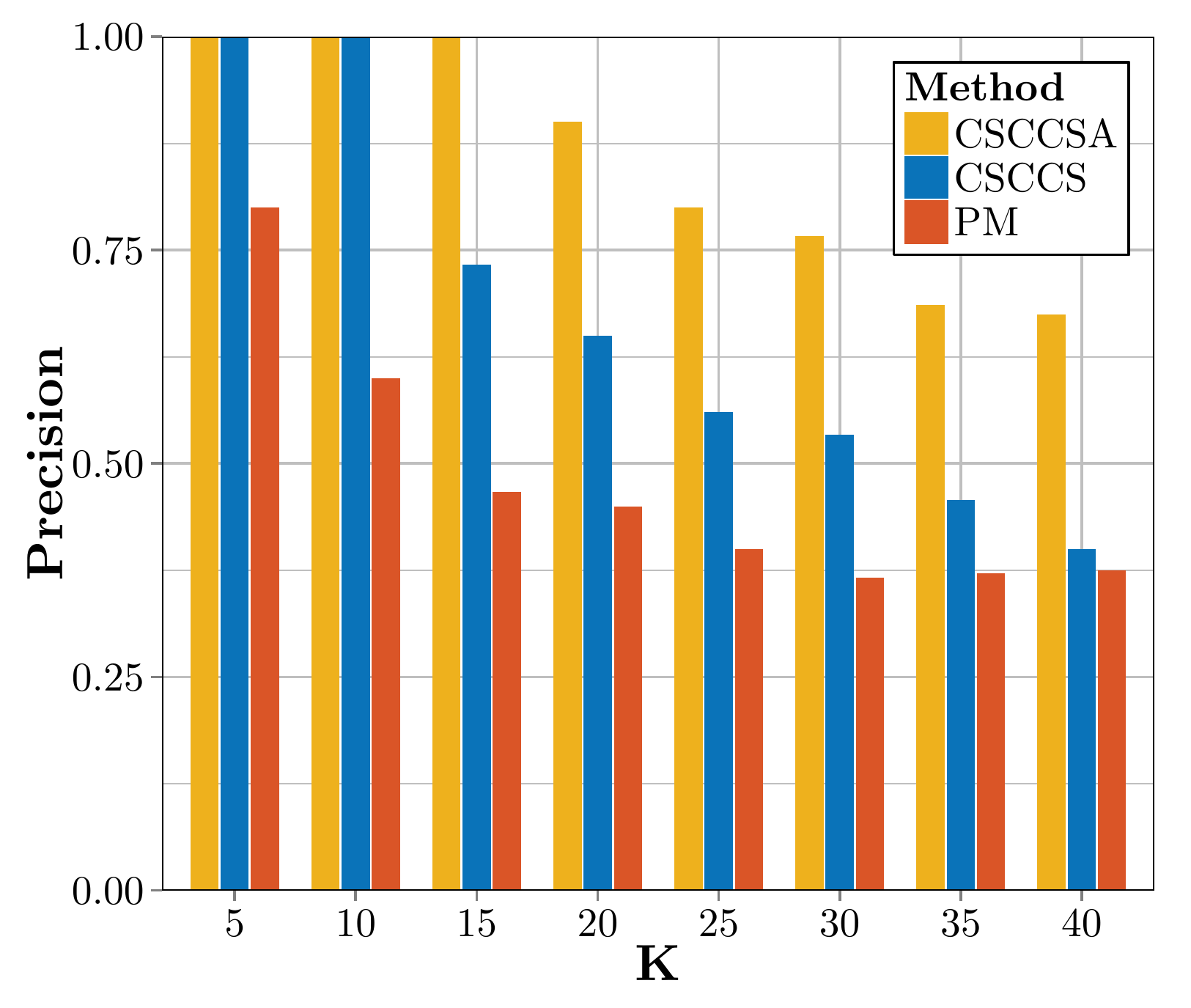}
\vspace{-4mm}
\caption{Precision at K of PM, CSCCS, and CSCCSA}
\label{fig:patk}\vspace{-4mm}
\end{figure}

The task of CDR from EHRs is somewhat analogous to web search. Specifically,  the algorithm should select only a few drugs that have interesting unexpected effects on the response: returning too many results makes it infeasible for human experts to evaluate the potential effect of the selected drugs. This is similar to users performing web search on a search engine, where typically only the quality of the results on the first page, or the first K results, matters. Based on this observation, an algorithm with a high precision-at-K value is desirable. Figure~\ref{fig:patk} shows the precision of each of the three algorithms at different positions (K) in the task of identifying drugs prescribed for lowering glucose. CSCCSA achieves the highest performance at all positions. CSCCS outperforms PM significantly at smaller K's, but the performances of the two algorithms are similar at larger K's. This is consistent with results in Table~\ref{tab:summary}, showing that CSCCSA is able to identify more prescribed drugs for lowering glucose than the other two methods.  Moreover, these drugs are at the very top of Table~\ref{tab:csccsa}. Therefore, precision-at-K provides evidence for CSCCSA's utility for CDR from EHRs.

\subsection{Drugs with Known Glucose Increasing/  Decreasing Effects}
From Tables~\ref{tab:pm}--\ref{tab:csccsa}, we notice that CSCCSA discovers the most number of drugs prescribed for lowering glucose among the three methods under consideration. This reaffirms our belief that CSCCSA is a promising method for CDR from EHR.  Furthermore, we also notice that drugs prescribed for increasing glucose  are reported in all but the table of CSCCSA. 

In Table~\ref{tab:pm}, sucrose is observed as a false positive using PM. Based on its count, this might be a spurious correlation in the data. This is even more probable when we consider the fact that the effect of sucrose on blood glucose level is short-term, and sucrose is not a drug that consistently enter patients' EHR for a long period of time. However, PM considers the glucose measurement records of the patients within two years before and after the first prescription of sucrose, during which many  stronger confounding factors could have occurred to alter the glucose level.

In Table~\ref{tab:csccs}, glucagon is identified. Glucagon is given to diabetic patients that take glucose-lowering drugs to avoid hypoglycemia. However, glucagon alone is not frequently administered. Therefore, in the data, we observe the co-occurrence of glucagon with various glucose-lowering drugs. While glucagon  alone increase blood glucose, combining with glucose-lowering drugs usually results in the decrease of blood sugar. On the other hand, we did not have enough data where glucagon is prescribed alone to observe the responses. Therefore, the algorithm will consider glucagon to have glucose-lowering effects since most of the time the occurrence of glucagon is accompanied by blood sugar decreasing medications. The algorithm might even consider it as a strong glucose-lowering drug because the actual glucose-lowering drugs are coded in various names in the EHR, hence dispersing the effect, while glucagon is coded only by a few different names.

\subsection{Confounding and Potential Drugs}
\label{sec:potential}
We now turn to the discussion of the drugs discovered by the three algorithms in Table~\ref{tab:pm}--\ref{tab:csccsa} that are not prescribed for glucose increasing/decreasing. We will make use of a list providing drugs that can influence blood glucose level available in \citep{DiabetesInControl} to aid our evaluation process.
\subsubsection{The Blessing and the Curse of Marginal Correlations}
According to \citep{DiabetesInControl}, in Table~\ref{tab:pm}, Actigall can cause blood glucose level to increase while amphotericin B can cause blood glucose level to decrease. An interesting drug that is also brought to our attention is buderprion SR. Buderprion SR is an antidepressant prescribed for the treatment of depressive disorder. For diabetic patients with depression, buderprion SR can help to alleviate their depressive symptom, making them in a better mood. This in turn has a positive effect on better controlling blood glucose level for longer period of time \citep{lustmanfactors2007}. PM is able to discover the blood glucose lowering effect of buderprion SR, even with a mere support of nine patients. The fact that PM considers the marginal correlation of each drug-indication pair independently makes it more likely to discover interesting drug-indication pairs with a weaker support. However, spurious correlations, especially those caused by the innocent bystander problem, are also more likely to be reported this way. 

Comparing the results from Table~\ref{tab:pm} with those from Table~\ref{tab:csccs} and Table~\ref{tab:csccsa} could justify our argument. In Table~\ref{tab:pm}, Habitrol is a nicotine patch, and Monistat, Voriconazole, amphotericin B, and Hibiclens are all used to treat fungal infection. Interestingly, fungal infection is a comorbidity of diabetes \citep{vazquezfungal1995,Skin74:online}, and smokers are also more inclined to be diabetic \citep{Smoki20:online}. On the other hand, we cannot find any drugs that are related to fungal infection or quitting smoke in Table~\ref{tab:csccs} and Table~\ref{tab:csccsa}. This comparison suggests that the aforementioned drugs in Table~\ref{tab:pm} generated by marginal association methods like PM might be innocent bystanders while a multiple regression approach such as CSCCS and CSCCSA might significantly help to alleviate this type of confounding issue.
\input{./support/table}

\subsubsection{Potential drugs found by CSCCS and CSCCSA}
In Table~\ref{tab:csccs},  a study \citep{vermesenalapril2003} indicates that enalapril helps to decrease the occurrence rate of diabetes in patients with chronic heart failure. Tricor might also have the potential to lower blood sugar level, based on the findings in \cite{damcifenofibrate2003} and \cite{balakumarfenofibrate2014}. Vitamin B12 is another interesting drug for consideration. In a rat model used by \cite{chowthe1957}, deficiency in vitamin B12 is linked to hyperglycemia. However, blood glucose level can be decreased by providing vitamin B12. A recent study suggests that diabetic patients under metformin might experience vitamin B12 deficiency \citep{tingrisk2006}. In a study on depressive patients, Zoloft, which is an antidepressant, is linked to the increase of insulin level after its prescription \citep{kesimthe2011}. Zestril, which is the brand name of lisinopril, is found to inhibit high blood sugar level in rats \citep{balakumarfenofibrate2014}. Captopril is also reported to improve daily glucose profile among non-insulin-dependent patients \citep{kodamaeffect1990}. However, hydralazine HCl is linked to glucose-increasing in a rat model, according to the findings in \cite{satoh1980hyperglycemic}. Nifedipine, verapamil HCl, and morphine sulfate can decrease blood sugar while captopril interacting with hydrochlorothiazide could cause high blood sugar, according to the list in \cite{DiabetesInControl}.  The potential glucose-lowering drugs discovered indicate that CSCCS is a reasonable method for the task of CDR.

In Table~\ref{tab:csccsa}, Pravachol is a member of a popular class of drugs called statins which are prescribed to lower cholesterol level.  Although the Food and Drug Administration (FDA) has added blood-glucose-increase warnings to all the drugs in the statin class \citep{fda}, Pravachol itself has been considered to have blood-glucose lowering effects \citep{freemanpravastatin2001,carterrisk2013}. The fact that CSCCSA can single out this particular drug from other statin class drug members indicates the potential of the algorithm to distinguish among similar drugs that have subtle  differences. Celexa has a mild but non-significant effect on FBG level reduction in a study with seventeen depressive patients \citep{amsterdamsafety2006}. Several cases of hypoglycemia linked to the use of Neurontin have also been reported \citep{schollsix2015}. Thiamine is reported to reduce the adverse effect of  hyperglycemia by inhibiting certain biological pathways \citep{vinhquocluongthe2012} and deficiency of thiamine is observed in diabetic patients \citep{page2011thiamine}. Cardura is found to reduce insulin resilience in a study on hypertensive patients with diabetes \citep{inukaiclinical2004}. According to \cite{DiabetesInControl}, Prozac can cause both high or lower blood sugar while diltiazem is linked to low blood glucose level.

\subsection{Experiments on Low-density Lipoprotein}
To demonstrate the potential of our methodology, we also apply our method to predict the numeric value of low-density lipoprotein (LDL). We first construct drug eras from drug prescription records with the approach proposed in Section~\ref{sec:bp}. We then run CSCCSA and generate a long list of about two hundred drugs. We report the top forty drugs from the list in Table~\ref{tab:ldl}. No confirmed false positives are discovered in the table while all the confirmed true positives are reported at the very top of the list. Some entries of hormone are discovered, which are linked to the decrease of LDL in drug/laboratory tests \citep{www.a73:online}. Interestingly, many entries of antibiotics are discovered, and all of them are classified as non-recurrent drugs by the algorithm in Section~\ref{sec:bp}. This is consistent with the clinical practice that antibiotics are usually not prescribed for long-term use. Some antibiotics have also been considered to manage cholesterol level, with literature support dating back to the 1950's \citep{samueltreatment1979, kesaniemiturnover1984, jenkinseffect2005}. The experimental results on LDL suggest that our algorithm is not fine-tuned to boost the performance on discovering drugs that control FBG level. Instead, it is readily applicable to other important numeric clinical measurements that might lead to interesting discoveries in drug repositioning.

\section{Discussion}
We have introduced the CSCCS model for the task of CDR using EHRs. To our best knowledge, the proposed model is the first of its kind to extensively leverage temporal ordering information from EHR to predict indications for multiple drugs at the same time. The CSCCS model extends the SCCS model that is popular in the ADR community to address a continuous response. As an initial effort, we evaluate our methodology on the task of discovering potential blood-sugar-lowering indications for a variety of drugs in a real world EHR. We develop a set of experimental evaluation methods specific to this problem in order to estimate the performance of our method. Our experimental results suggest that CSCCS can not only discover existing indications  but is also able to identify potentially new use of drugs. We hence believe that CSCCS is a promising model to aid the knowledge discovery process in CDR.

Future applications and extensions of the CSCCS model are exciting. To begin with, CSCCS can be applied to a broad variety of numeric responses such as  blood pressure level, cholesterol level, or body weight, to name a few. Therefore, potentially new indications of drugs to control the aforementioned important physical measurements can be examined in the same paradigm. Furthermore, many other sources of patient information, such as demographic information, diagnosis codes, other type of lab measurements, as well as interactions among all these information sources can be taken into consideration to facilitate the prediction of the physical measurement level. Last but not least,  although the proposed CSCCS model in its simpliest form is a linear model, the history of SCCS model development \citep{madiganself} could help guide on future development of CSCCS model. More complicated models can be derived from its simpler counterparts for better predictive performance in more specific and refined applications.

We can learn a lot of lessons from the task of ADR discovery from EHRs to perform the task of CDR from EHRs. For one thing, a ground truth set for CDR algorithm evaluation using EHRs is desired and calls for multidisciplinary collaborations from  experts with different specialties. Furthermore, the popularity of disproportionality analysis and self-controlled methods in the ADR discovery community suggestes the utility of analogous methods for CDR.

\subsection*{Acknowledgments}
The authors would like to gratefully acknowledge the anonymous reviewers for their reviewing efforts and invaluable suggestions.

\small{\bibliography{kuang2016}}
\end{document}

%% file: support/ehr.pdf_tex
\begingroup%
  \makeatletter%
  \providecommand\color[2][]{%
    \errmessage{(Inkscape) Color is used for the text in Inkscape, but the package 'color.sty' is not loaded}%
    \renewcommand\color[2][]{}%
  }%
  \providecommand\transparent[1]{%
    \errmessage{(Inkscape) Transparency is used (non-zero) for the text in Inkscape, but the package 'transparent.sty' is not loaded}%
    \renewcommand\transparent[1]{}%
  }%
  \providecommand\rotatebox[2]{#2}%
  \ifx\svgwidth\undefined%
    \setlength{\unitlength}{198.77145395bp}%
    \ifx\svgscale\undefined%
      \relax%
    \else%
      \setlength{\unitlength}{\unitlength * \real{\svgscale}}%
    \fi%
  \else%
    \setlength{\unitlength}{\svgwidth}%
  \fi%
  \global\let\svgwidth\undefined%
  \global\let\svgscale\undefined%
  \makeatother%
  \begin{picture}(1,0.7504415)%
    \put(-0.31323895,0.51890363){\color[rgb]{0,0,0}\makebox(0,0)[lb]{\smash{}}}%
    \put(0.42497862,0.72330214){\color[rgb]{0,0,0}\makebox(0,0)[lb]{\smash{Person 1}}}%
    \put(0,0){\includegraphics[width=\unitlength,page=1]{ehr.pdf}}%
    \put(0.86629123,0.4104077){\color[rgb]{0,0,0}\makebox(0,0)[lb]{\smash{567}}}%
    \put(0,0){\includegraphics[width=\unitlength,page=2]{ehr.pdf}}%
    \put(0.06602679,0.58703663){\color[rgb]{0,0,0}\makebox(0,0)[lb]{\smash{Glucose=150}}}%
    \put(0,0){\includegraphics[width=\unitlength,page=3]{ehr.pdf}}%
    \put(0.32309527,0.66934619){\color[rgb]{0,0,0}\makebox(0,0)[lb]{\smash{Glucose=200}}}%
    \put(0,0){\includegraphics[width=\unitlength,page=4]{ehr.pdf}}%
    \put(0.7471287,0.55047316){\color[rgb]{0,0,0}\makebox(0,0)[lb]{\smash{Glucose=100}}}%
    \put(0,0){\includegraphics[width=\unitlength,page=5]{ehr.pdf}}%
    \put(0.23725228,0.47893349){\color[rgb]{0,0,0}\makebox(0,0)[lb]{\smash{Drug 1}}}%
    \put(0.13181973,0.4227366){\color[rgb]{0,0,0}\makebox(0,0)[lb]{\smash{1}}}%
    \put(0.42502775,0.3328689){\color[rgb]{0,0,0}\makebox(0,0)[lb]{\smash{Person 2}}}%
    \put(0,0){\includegraphics[width=\unitlength,page=6]{ehr.pdf}}%
    \put(0.56183439,0.08867073){\color[rgb]{0,0,0}\makebox(0,0)[lb]{\smash{Drug 1}}}%
    \put(0,0){\includegraphics[width=\unitlength,page=7]{ehr.pdf}}%
    \put(0.16977053,0.22075172){\color[rgb]{0,0,0}\makebox(0,0)[lb]{\smash{Glucose=160}}}%
    \put(0,0){\includegraphics[width=\unitlength,page=8]{ehr.pdf}}%
    \put(0.01631729,0.25476461){\color[rgb]{0,0,0}\makebox(0,0)[lb]{\smash{Glucose=180}}}%
    \put(0,0){\includegraphics[width=\unitlength,page=9]{ehr.pdf}}%
    \put(0.73013076,0.14394103){\color[rgb]{0,0,0}\makebox(0,0)[lb]{\smash{Glucose=80}}}%
    \put(0,0){\includegraphics[width=\unitlength,page=10]{ehr.pdf}}%
    \put(0.91992811,0.01606077){\color[rgb]{0,0,0}\makebox(0,0)[lb]{\smash{643}}}%
    \put(0,0){\includegraphics[width=\unitlength,page=11]{ehr.pdf}}%
    \put(0.32578786,0.11079381){\color[rgb]{0,0,0}\makebox(0,0)[lb]{\smash{Drug 2}}}%
    \put(0.04056659,0.03034873){\color[rgb]{0,0,0}\makebox(0,0)[lb]{\smash{1}}}%
  \end{picture}%
\endgroup%

%% file: support/table.tex
\begin{table*}[t!]
\centering
\parbox[t]{.49\textwidth}{%
\centering 
\caption{Top forty drugs: PM-Glucose}
\label{tab:pm}\vspace{-0.25cm}
\input{./support/pmTab.tex}}
\parbox[t]{.49\textwidth}{
\centering 
\caption{Top forty drugs: CSCCS-Glucose}
\label{tab:csccs}\vspace{-0.25cm}
\input{./support/csccsTab.tex}}
\\ 
\vspace{0.2cm}
\parbox[t]{.49\textwidth}{
\centering 
\caption{Top forty drugs: CSCCSA-Glucose}
\label{tab:csccsa}\vspace{-0.25cm}
\input{./support/csccsaTab.tex}}
\parbox[t]{.49\textwidth}{
\centering 
\caption{Top forty drugs: CSCCSA-LDL}
\label{tab:ldl}\vspace{-0.25cm}
\input{./support/csccsaLdlTab.tex}}
\end{table*}

%% file: support/pmTab.tex
\scalebox{0.6}{
\begin{tabular}{c c C{7cm} C{2cm} c}
  \hline
INDX & CODE & DRUG NAME & SCORE & COUNT \\ 
  \hline
\rowcolor{Green}
1 & 5226 & LANTUS & -41.672 &   34 \\ 
\rowcolor{Green}
  2 & 6646 & NOVOFINE 31 & -38.709 &   33 \\ 
\rowcolor{Green}
  3 & 5789 & METFORMIN HYDROCHLORIDE & -38.623 &   10 \\ 
  4 & 5806 & METHENAM/MBLU/BA/SAL/ATROP/HYO & -36.710 &   10 \\ 
\rowcolor{Green}
  5 & 4811 & INSULIN NPH & -34.573 &   23 \\ 
\rowcolor{Green}
  6 & 6652 & NOVOLOG & -29.895 &   54 \\ 
  7 & 4336 & HABITROL & -29.871 &   16 \\ 
  8 & 6044 & MONISTAT & -29.721 &   14 \\ 
  9 & 9080 & SURFAK & -29.655 &   14 \\ 
\rowcolor{Green}
  10 & 9155 & SYRNG W-NDL DISP INSUL 0.333ML & -29.439 &   30 \\ 
\rowcolor{Green}
  11 & 4500 & HUMULIN & -29.186 &   36 \\ 
  12 & 9008 & SUGAR SUBSTITUTE & -28.971 &   10 \\ 
  13 & 10176 & VORICONAZOLE & -28.538 &   10 \\ 
  14 & 1305 & BUDEPRION SR & -27.444 &    9 \\ 
  15 & 8450 & ROXICODONE & -27.428 &   12 \\ 
  16 & 9534 & TRANDATE & -25.978 &    8 \\ 
\rowcolor{Green}
  17 & 4802 & INSULIN & -24.507 &  697 \\ 
  18 & 3849 & FLURBIPROFEN SODIUM & -24.403 &   11 \\ 
\rowcolor{Green}
  19 & 8316 & REZULIN & -24.287 &  135 \\ 
  20 & 5257 & LENALIDOMIDE & -22.875 &    8 \\ 
\rowcolor{Green}
  21 & 4485 & HUMALOG & -22.852 &   67 \\ 
  22 & 1389 & CAL & -22.817 &   61 \\ 
  23 &  144 & ACTIGALL & -22.237 &   36 \\ 
\rowcolor{Red}
  24 & 8998 & SUCROSE & -22.125 &   18 \\ 
  25 & 3843 & FLUPHENAZINE HCL & -22.094 &    8 \\ 
  26 & 3682 & FERROUS FUMARATE & -21.225 &   10 \\ 
  27 & 9104 & SYMLINPEN 120 & -20.333 &   12 \\ 
  28 & 1868 & CHLORAMBUCIL & -20.268 &   14 \\ 
\rowcolor{Green}
  29 & 4171 & GLUCOTROL XL & -19.719 &  828 \\ 
  30 &  504 & AMPHOTERICIN B & -19.672 &   24 \\ 
  31 & 3778 & FLEXOR & -19.287 &   14 \\ 
\rowcolor{Green}
  32 & 8241 & REGULAR INSULIN & -19.205 &   39 \\ 
\rowcolor{Green}
  33 &  824 & AVANDIA & -19.140 &  487 \\ 
  34 & 5783 & METAPROTERENOL & -18.920 &   10 \\ 
  35 & 4434 & HIBICLENS & -18.863 &   10 \\ 
  36 & 5815 & METH/ME BLUE/BA/PHENY/ATP/HYOS & -18.727 &   11 \\ 
\rowcolor{Green}
  37 & 5010 & JANUVIA & -18.716 &   11 \\ 
\rowcolor{Green}
  38 & 4813 & INSULIN NPL/INSULIN LISPRO & -18.515 &  126 \\ 
  39 & 4595 & HYDROMORPHONE & -18.470 &   17 \\ 
  40 & 7626 & POLYMYXIN B SULFATE MICRONIZED & -18.456 &   11 \\ 
   \hline
\end{tabular}
}

%% file: support/csccsTab.tex
\scalebox{0.6}{
\begin{tabular}{c c C{7cm} C{2cm} c}
  \hline
INDX & CODE & DRUG NAME & SCORE & COUNT \\ 
  \hline
\rowcolor{Green}
1 & 7470 & PIOGLITAZONE HCL & -13.502 & 3075 \\ 
\rowcolor{Green}
  2 & 8437 & ROSIGLITAZONE MALEATE & -13.465 & 1019 \\ 
\rowcolor{Green}
  3 & 6656 & NPH HUMAN INSULIN ISOPHANE & -10.963 & 2874 \\ 
\rowcolor{Green}
  4 & 4497 & HUM INSULIN NPH/REG INSULIN HM & -10.869 & 1829 \\ 
\rowcolor{Green}
  5 &  160 & ACTOS & -7.665 & 1125 \\ 
\rowcolor{Green}
  6 &  824 & AVANDIA & -7.543 & 1239 \\ 
\rowcolor{Green}
  7 & 4837 & INSULN ASP PRT/INSULIN ASPART & -7.067 &  258 \\ 
\rowcolor{Green}
  8 & 4806 & INSULIN GLARGINE HUM.REC.ANLOG & -5.571 & 4213 \\ 
\rowcolor{Green}
  9 & 9152 & SYRING W-NDL DISP INSUL 0.5ML & -5.301 & 4186 \\ 
\rowcolor{Green}
  10 & 8316 & REZULIN & -3.611 &  444 \\ 
  11 & 3227 & ENALAPRIL & -3.218 & 1103 \\ 
\rowcolor{Green}
  12 & 6382 & NEEDLES INSULIN DISPOSABLE & -3.148 & 2827 \\ 
  13 & 4970 & ISOSORBIDE DINITRATE & -3.122 & 1220 \\ 
  14 & 9623 & TRICOR & -3.119 &  821 \\ 
  15 & 3686 & FERROUS SULFATE & -2.898 & 4820 \\ 
  16 & 1760 & CELEXA & -2.887 & 1473 \\ 
\rowcolor{Green}
  17 & 4802 & INSULIN & -2.806 & 1526 \\ 
\rowcolor{Red}
  18 & 4118 & GLUCAGON HUMAN RECOMBINANT & -2.722 & 1639 \\ 
\rowcolor{Green}
  19 & 5786 & METFORMIN & -2.625 & 3838 \\ 
  20 & 7731 & PRAVACHOL & -2.458 & 1700 \\ 
  21 & 2512 & DARBEPOETIN ALFA IN ALBUMN SOL & -2.359 &  426 \\ 
  22 & 6210 & MYCOPHENOLATE MOFETIL & -2.253 &  724 \\ 
  23 & 2830 & DILTIAZEM & -2.216 & 1021 \\ 
  24 & 5636 & MAVIK & -2.150 & 2242 \\ 
\rowcolor{Green}
  25 & 4132 & GLUCOPHAGE & -2.133 & 6736 \\ 
  26 & 4525 & HYDRALAZINE HCL & -2.095 &  792 \\ 
\rowcolor{Green}
  27 & 4106 & GLIMEPIRIDE & -2.034 & 3384 \\ 
  28 & 7129 & PAXIL & -2.033 & 2021 \\ 
  29 & 2426 & CYANOCOBALAMIN (VITAMIN B-12) & -1.992 & 4080 \\ 
\rowcolor{Green}
  30 & 4833 & INSULIN ZINC HUMAN REC & -1.945 &  116 \\ 
  31 & 10392 & ZOLOFT & -1.926 & 2417 \\ 
  32 & 6069 & MORPHINE SULFATE & -1.889 &  899 \\ 
  33 & 10333 & ZESTRIL & -1.787 & 2032 \\ 
  34 & 1216 & BLOOD SUGAR DIAGNOSTIC & -1.665 & 19832 \\ 
  35 & 10199 & WARFARIN SODIUM & -1.632 & 9223 \\ 
  36 & 3937 & FOSINOPRIL SODIUM & -1.540 & 2660 \\ 
  37 & 6499 & NIFEDIPINE & -1.524 & 1472 \\ 
  38 & 1003 & BENAZEPRIL HCL & -1.462 & 1586 \\ 
  39 & 9994 & VERAPAMIL HCL & -1.433 & 1856 \\ 
  40 & 1573 & CAPTOPRIL & -1.418 & 1989 \\ 
   \hline
\end{tabular}
}

%% file: support/csccsaTab.tex
\scalebox{0.6}{
\begin{tabular}{c c C{7cm} C{2cm} c}
  \hline
INDX & CODE & DRUG NAME & SCORE & COUNT \\ 
  \hline
\rowcolor{Green}
1 & 4485 & HUMALOG & -11.786 &  124 \\ 
\rowcolor{Green}
  2 & 7470 & PIOGLITAZONE HCL & -10.220 & 3075 \\ 
\rowcolor{Green}
  3 & 8437 & ROSIGLITAZONE MALEATE & -9.731 & 1019 \\ 
\rowcolor{Green}
  4 & 4837 & INSULN ASP PRT/INSULIN ASPART & -9.658 &  258 \\ 
\rowcolor{Green}
  5 & 6382 & NEEDLES INSULIN DISPOSABLE & -9.464 & 2827 \\ 
\rowcolor{Green}
  6 & 4171 & GLUCOTROL XL & -8.117 & 2853 \\ 
\rowcolor{Green}
  7 & 4106 & GLIMEPIRIDE & -7.940 & 3384 \\ 
\rowcolor{Green}
  8 &  160 & ACTOS & -7.721 & 1125 \\ 
\rowcolor{Green}
  9 &  824 & AVANDIA & -6.802 & 1239 \\ 
\rowcolor{Green}
  10 & 9152 & SYRING W-NDL DISP INSUL 0.5ML & -6.623 & 4186 \\ 
\rowcolor{Green}
  11 & 4132 & GLUCOPHAGE & -6.322 & 6736 \\ 
\rowcolor{Green}
  12 & 4184 & GLYBURIDE & -6.021 & 8879 \\ 
\rowcolor{Green}
  13 & 4170 & GLUCOTROL & -5.721 & 1259 \\ 
\rowcolor{Green}
  14 & 4208 & GLYNASE & -5.670 &  591 \\ 
\rowcolor{Green}
  15 &  416 & AMARYL & -5.599 & 2240 \\ 
\rowcolor{Green}
  16 & 4107 & GLIPIZIDE & -5.563 & 9993 \\ 
  17 &  844 & AXID & -4.682 &  189 \\ 
  18 & 2830 & DILTIAZEM & -4.297 & 1021 \\ 
\rowcolor{Green}
  19 & 4806 & INSULIN GLARGINE HUM.REC.ANLOG & -4.175 & 4213 \\ 
\rowcolor{Green}
  20 & 5787 & METFORMIN HCL & -4.147 & 19584 \\ 
  21 & 2824 & DILAUDID & -4.076 &   39 \\ 
\rowcolor{Green}
  22 & 5786 & METFORMIN & -3.890 & 3838 \\ 
  23 & 7731 & PRAVACHOL & -3.532 & 1700 \\ 
  24 & 1760 & CELEXA & -3.517 & 1473 \\ 
\rowcolor{Green}
  25 & 4497 & HUM INSULIN NPH/REG INSULIN HM & -3.501 & 1829 \\ 
  26 & 9889 & URSODIOL & -3.132 &  376 \\ 
\rowcolor{Green}
  27 & 4813 & INSULIN NPL/INSULIN LISPRO & -2.972 &  623 \\ 
\rowcolor{Green}
  28 & 4133 & GLUCOPHAGE XR & -2.845 &  765 \\ 
  29 & 6445 & NEURONTIN & -2.615 & 1418 \\ 
\rowcolor{Green}
  30 & 6656 & NPH HUMAN INSULIN ISOPHANE & -2.500 & 2874 \\ 
  31 & 9379 & THIAMINE HCL & -2.383 &  341 \\ 
  32 & 1636 & CARDURA & -2.198 & 1079 \\ 
  33 & 1218 & BLOOD SUGAR DIAGNOSTIC DRUM & -2.073 & 2593 \\ 
  34 & 8025 & PROZAC & -2.037 & 1525 \\ 
\rowcolor{Green}
  35 & 8316 & REZULIN & -1.895 &  444 \\ 
\rowcolor{Green}
  36 & 9136 & SYRINGE \& NEEDLE INSULIN 1 ML & -1.885 & 3542 \\ 
\rowcolor{Green}
  37 & 4802 & INSULIN & -1.812 & 1526 \\ 
  38 & 7674 & POTASSIUM CHLORIDE & -1.779 & 9842 \\ 
\rowcolor{Green}
  39 & 4804 & INSULIN ASPART & -1.752 & 2476 \\ 
  40 & 1200 & BLOOD-GLUCOSE METER & -1.719 & 5289 \\ 
   \hline
\end{tabular}
}

%% file: support/csccsaLdlTab.tex
\scalebox{0.6}{
\begin{tabular}{c c C{7cm} C{2cm} c}
  \hline
INDX & CODE & DRUG NAME & SCORE & COUNT \\ 
  \hline
\rowcolor{Green}
1 & 8444 & ROSUVASTATIN CALCIUM & -17.052 & 27122 \\ 
\rowcolor{Green}
  2 & 5368 & LIPITOR & -16.908 & 118468 \\ 
\rowcolor{Green}
  3 & 2395 & CRESTOR & -16.234 & 3535 \\ 
\rowcolor{Green}
  4 & 8720 & SIMVASTATIN & -15.790 & 206064 \\ 
\rowcolor{Green}
  5 & 3584 & EZETIMIBE/SIMVASTATIN & -14.721 & 19396 \\ 
\rowcolor{Green}
  6 &  790 & ATORVASTATIN CALCIUM & -13.982 & 151106 \\ 
\rowcolor{Green}
  7 &  941 & BAYCOL & -12.924 & 1236 \\ 
\rowcolor{Green}
  8 & 10383 & ZOCOR & -11.451 & 26514 \\ 
\rowcolor{Green}
  9 & 10186 & VYTORIN & -9.877 & 9047 \\ 
\rowcolor{Green}
  10 & 5487 & LOVASTATIN & -9.238 & 45286 \\ 
\rowcolor{Green}
  11 & 3583 & EZETIMIBE & -8.093 & 32595 \\ 
\rowcolor{Green}
  12 & 7731 & PRAVACHOL & -6.729 & 16525 \\ 
\rowcolor{Green}
  13 & 10336 & ZETIA & -6.678 & 6623 \\ 
\rowcolor{Green}
  14 & 7733 & PRAVASTATIN SODIUM & -6.638 & 33708 \\ 
\rowcolor{Green}
  15 & 5261 & LESCOL XL & -6.358 &  873 \\ 
  16 & 9183 & TAMOXIFEN CITRATE & -4.777 & 3095 \\ 
\rowcolor{Green}
  17 & 5893 & MEVACOR & -4.172 & 4205 \\ 
  18 & 2175 & COLACE & -4.016 & 4349 \\ 
  19 & 9182 & TAMOXIFEN & -3.764 & 2048 \\ 
\rowcolor{Green}
  20 & 5260 & LESCOL & -3.716 & 6251 \\ 
\rowcolor{Green}
  21 &  475 & AMLODIPINE/ATORVASTATIN & -2.779 & 1272 \\ 
  22 &  494 & AMOXICILLIN/POTASSIUM CLAV & -2.495 & 4186 \\ 
  23 & 2110 & CLOPIDOGREL BISULFATE & -2.271 & 50059 \\ 
  24 & 4616 & HYDROXYCHLOROQUINE SULFATE & -2.240 & 5888 \\ 
  25 & 5281 & LEVAQUIN & -2.194 & 1464 \\ 
  26 & 3471 & ESTROGEN CON/M-PROGEST ACET & -1.929 & 5896 \\ 
  27 & 7496 & PLAVIX & -1.471 & 14220 \\ 
  28 & 8225 & RED YEAST RICE & -1.345 & 5468 \\ 
  29 & 3746 & FLAGYL & -1.169 &  278 \\ 
  30 & 6540 & NITROGLYCERIN & -1.103 & 94747 \\ 
  31 & 2959 & DOCUSATE SODIUM & -1.084 & 32872 \\ 
  32 & 3475 & ESTROGENS CONJUGATED & -1.033 & 22480 \\ 
  33 & 3686 & FERROUS SULFATE & -0.990 & 32496 \\ 
  34 & 7768 & PREMARIN & -0.969 & 5513 \\ 
  35 &  865 & AZITHROMYCIN & -0.959 & 9861 \\ 
  36 & 2811 & DIGOXIN & -0.908 & 31353 \\ 
  37 & 4132 & GLUCOPHAGE & -0.779 & 14764 \\ 
  38 &  493 & AMOXICILLIN & -0.715 & 11214 \\ 
  39 & 1985 & CIPROFLOXACIN & -0.651 &  989 \\ 
  40 & 9946 & VARENICLINE TARTRATE & -0.636 & 10794 \\ 
   \hline
\end{tabular}
}